\documentclass[a4paper,fleqn]{cas-dc}

\usepackage{soul, color, xcolor}
\usepackage{color,array,amsmath}
\usepackage{graphicx}
\usepackage{overpic}
\usepackage{graphicx}
\usepackage{multirow}
\usepackage{booktabs}
\usepackage{amsmath}
\usepackage{float}  
\usepackage{lipsum}
\usepackage{lineno,hyperref}
\usepackage[round]{natbib}
\usepackage[ruled]{algorithm2e}
\usepackage{algpseudocode}
\usepackage{multirow}
\usepackage{booktabs}
\usepackage{color, soul} 
\usepackage {hyperref}
\usepackage{wrapfig}
\usepackage{orcidlink}



\begin{document}
\let\printorcid\relax
\title {\Large Mode-conditioned music learning and composition: a spiking neural network inspired by neuroscience and psychology}
\shorttitle {SNN-based mode-conditioned music learning and composition}
\shortauthors{Q.Liang, Y.Zeng and MHR.Tang}

\author[1]{Qian Liang \orcidlink{0000-0003-2424-7086}}

\author{Yi Zeng,\orcidlink{0000-0002-9595-9091} $^{a,b,c,*}$ }
\cortext[cor1]{Corresponding author: yi.zeng@ia.ac.cn, 100190}

\author[2]{Menghaoran Tang}

\address[1]{Brain-inspired Cognitive Intelligence Lab, Institute of Automation, Chinese Academy of Sciences, Beijing, China.}
\address[2]{University of Chinese Academy of Sciences, Beijing, China.}
\address[3]{Key Laboratory of Brain Cognition and Brain-inspired Intelligence Technology, Chinese Academy of Sciences, Shanghai, China.}

\begin{abstract}
Musical mode is one of the most critical element that establishes the framework of pitch organization and determines the harmonic relationships. Previous works often use the simplistic and rigid alignment method, and overlook the diversity of modes. However, in contrast to AI models, humans possess cognitive mechanisms for perceiving the various modes and keys. In this paper, we propose a spiking neural network inspired by brain mechanisms and psychological theories to represent musical modes and keys, ultimately generating musical pieces that incorporate tonality features. Specifically, the contributions are detailed as follows: 1) The model is designed with multiple collaborated subsystems inspired by the structures and functions of corresponding brain regions; 2)We incorporate mechanisms for neural circuit evolutionary learning that enable the network to learn and generate mode-related features in music, reflecting the cognitive processes involved in human music perception. 3)The results demonstrate that the proposed model shows a connection framework closely similar to the Krumhansl-Schmuckler model, which is one of the most significant key perception models in the music psychology domain. 4) Experiments show that the model can generate music pieces with characteristics of the given modes and keys. Additionally, the quantitative assessments of generated pieces reveals  that the generating music pieces have both tonality characteristics and the melodic adaptability needed to generate diverse and musical content. By combining insights from neuroscience, psychology, and music theory with advanced neural network architectures, our research aims to create a system that not only learns and generates music but also bridges the gap between human cognition and artificial intelligence.
\end{abstract}


\maketitle
\section{Introduction}\label{sec:1}
Music learning and generation have long fascinated researchers across various disciplines, including neuroscience, psychology, and artificial intelligence. In particular, the artificial intelligence field has witnessed significant advancements in recent years, driven by the integration of deep learning techniques, large-scale models, and novel neural network architectures~\cite{briot2020deep,Hernandez2022}. With the emergence of OpenAI's Sora~\mbox{\cite{sora}}, the techniques of music learning and generation seem so powerful that they are capable of producing compositions that rival human-created music in complexity and emotional depth. However, despite these impressive advancements, the essential challenge also remains: current techniques lack the ability to understand and generate music on a cognitive level. This challenge gives rise to two key problems: 1) current computational model do not have the real ability of music perception and understanding, meaning they cannot interpret or comprehend music in the way humans do, 2) results of music learning and generating consequently fail to capture the depth, nuance, and creativity process inherent to human composers, finally limiting the consistency and musicality of generated results. 

Music is intrinsic to human nature, and researchers of cognitive neuroscience have shown that it is a so complex cognitive activity requiring multiple brain areas to collaborate~\cite{koelsch2012brain}. The perception and creation of music involve intricate processes that engage various regions of the brain. The primary auditory cortex (A1) plays a crucial role in sound processing, encoding the real sound signals to neural electrophysiological signals~\cite{hodges2019oxford}. The memory system is responsible for storing the sequential information of ordered notes, which is fundamental for recognizing familiar melodies, recalling harmonic progressions, and understanding complex musical forms. Prefrontal cortex, which is the most important region to represent concept and knowledge of the world, plays a crucial role in guiding learning and creating behaviors of human~\cite{bi2021dual}. Additionally, music has a profound emotional impact, with the emotion pathway, including the amygdala and orbitofrontal cortex playing central roles in encoding and generating emotional responses to music~\cite{hodges2019oxford}. The integration of these systems enables humans to experience and create music in a deeply meaningful and expressive way. Understanding these neural mechanisms provides a foundation for developing models that aim to make AI models more cognitive inspired, further improving their ability to generate music with enhanced emotional depth, coherence, and human-like creativity.

Mode theory and key play a central role in music composition, serving as the foundation for establishing tonal centers and guiding the melodic and harmonic structure of a piece. Together, they provide a coherent framework that shapes the emotional character and direction of music. The interplay between mode and key is crucial in creating process, influencing how listeners perceive mood and movement in a composition. Despite its importance, this fundamental theory is often overlooked by current AI music generation models. Most existing models adopt a simplified approach by aligning all keys to either the C major or minor modes, disregarding the subtle tonal colors and unique relationships between notes in different modes. However, researchers in music psychology domain have found that humans possess innate key perception mechanisms, enabling them to distinguish and interpret different keys and modes based on tonal hierarchies. To explain this cognitive process, one of the most well-known models, the \textbf{Krumhansl-Schmuckler Model(KS)}, have been proposed. This model demonstrates how listeners perceive tonal centers and assign hierarchical importance to pitch classes within a key, providing a quantitative framework for understanding human key perception. Despite these advances in cognitive music theory, current AI models rarely incorporate such insights, limiting their ability to generate music that reflects the nuanced key perception and tonal complexity inherent in human compositions. Consequently, there remains a significant gap between AI-generated music and the complexity of human creativity.

Inspired by the mechanisms of human music cognition, and addressing the limitations of current techniques in capturing the cognitive depth of music perception and composition, this paper proposes a multi-region collaborated brain-inspired spiking neural network that learns and generates multi-track music by integrating key principles from brain mechanisms, music psychology findings, and Western mode and key theories. The contributions of this paper are as follows:
\begin{itemize}
\item 
We build a multi-area collaborative model\footnote{We released our code in \url{https://github.com/BrainCog-X/Brain-Cog/tree/main/examples/Knowledge\_Representation\_and\_Reasoning/musicMemory/task}} based on a Spiking Neural Network (SNN) on the open platform \textbf{BrainCog}~\cite{BrainCog}, inspired by neuroscience findings, to perceive, learn, and create four-part music based on the symbolic music form. We employ the Izhikevich neuron model within the SNN and utilize synaptic creation along with the Spike-Timing-Dependent Plasticity (STDP) learning rule to facilitate dynamic circuit evolution, which enhances the network's adaptability. 
\item 
The model incorporates Western mode theory, inspired by the mechanisms of representation of prior knowledge of prefrontal cortex, building a hierarchical subsystem to represent and learn modes, keys and the relationship between them. By integrating mode theory, our model aims to capture tonal characteristics and exhibit a connection framework similar to the well-known \textbf{Krumhansl-Schmuckler} pitch profile model. For understanding the model clearly, the musical terms used in this paper are listed in Table.\ref{tab:terms}
\item 
We build an evaluation framework including both model evaluation and generation quantitative assessments. The results show that the architecture of our proposed model is consistent with the Krumhansl-Schmuckler psychological model, and the qualitative assessment of the generated samples demonstrates that the generated pieces effectively capture the intended tonal characteristics of different modes and keys, further supporting the efficacy of our approach in bridging the gap between artificial intelligence and human-like musicality.
\end{itemize}

\section{Related Works}\label{sec:2}
\subsection{Traditional AI}
Symbolic music learning primarily focuses on music analysis. Current popular AI models on for symbolic music learning include Recurrent Neural Networks (RNNs), Variational AutoEncoders (VAEs), the Generative Adversarial Networks (GANs),  Transformers and etc. Previous Works using RNNs and LSTMs targeted melody generation~\cite{todd1989connectionist,ji2023survey, CONCERT1994, briot2020deep},  harmonic~\cite{eck2002first,magenta2016,boulanger2012modeling}, chord accompaniment~\cite{eck2002first}, polyphonic music~\cite{boulanger2012modeling} and stylistic variation~\cite{eck2002first, hadjeres2017deepbach}. VAEs and GANs have also been employed to improve polyphony~\cite{PianoTree}, style adaptation~\cite{JazzGAN}, and multi-track generation~\cite{roberts2018hierarchical, VAE2022, MidiNet, dong2018musegan}, Recently, Transformers-based models recently have gained popularity in the music generation domain~\cite{huang2020pop}. Then large language models (LLMs) based works ~\cite{MuseNet, Zhao2024AdversarialMidiBERTSM, xu2024generating} have further enhanced symbolic music generation. However, these computational models predominantly rely on large datasets and parameter-heavy architectures, overlooking the central role of mode and key theories in music composition. Furthermore, they diverge significantly from the intuitive, explainable, and cognitively driven learning and creative processes exhibited by human composers, resulting in outputs that lack the depth and coherence of human-generated music.

\subsection{Neuroscience and Psychology}
Studies have shown that the primary auditory cortex(A1) has a topographical structure that maps the cochlea. The organization implies that neural populations in different regions response to different frequencies of sound~\cite{vuust2022music,purves2001,sankaran2024encoding,mcdermott2008music}.  Theses findings clearly demonstrated that the encoding manner of the sound in brain. However, through there are no clear and consistent findings regarding  how the brain encodes and process temporal information in music perception, some researches have found that a significant number of cells in the medial premotor cortex (MPC) are sensitive to a range of signal durations, with their preferred durations spanning intervals from several hundred milliseconds~\cite{merchant2013neural,gupta2014processing}, and "Time Cells" ~\cite{macdonald2011hippocampal} may encode moments between temporal events and locations. These findings may be helpful for music time perception. Besides, one significant area of exploration is the role of memory in music cognition. Memory allows for the recognition of familiar melodies, the anticipation of musical patterns, and the contextual understanding of musical pieces. This involves both short-term memory, which helps in processing and interpreting ongoing musical input, and long-term memory, which stores musical knowledge and experiences accumulated over time. Researchers have found that, hippocampus, medial temporal lobe(MTL) as well as prefrontal cortex(PFC) are deeply involved in temporal context memory establishment~\cite{Shannon2011Medial,Mcandrews1991The,Meier2013Implicit,Jenkins2010Prefrontal}. Prior knowledge also exerts critical influence on memory and is one of the prominent factor guiding the learning behaviors across human lifespan. Studies have highlighted the significant roles of the medial prefrontal cortex (mPFC) and the hippocampus (HC) in the formation of prior knowledge and its utilization during the successful encoding, consolidation, and retrieval of memories~\cite{brod2013influence}.

Additionally, the \textbf{Krumhansl-Schmuckler Model}~\cite{krumhansl2001cognitive,temperley1999s}, which is one of the most important model in music psychology domain, developed the idea of tonal hierarchies through empirical experiments, where listeners rated the importance or fit of each note within a given key. For each major and minor key, Krumhansl constructed a key profile, which represents the relative importance or frequency of the 12 pitch classes (C, C\#, D, D\#, etc.). Each profile consists of values based on the experimental data from listeners. The model compares the pitch-class distribution from the musical input to each of the 24 key profiles (12 major and 12 minor). It uses correlation to determine how well the pitch-class distribution matches each key profile. The key that yields the highest correlation is considered the most likely key of the piece. This model plays a significant role in understanding tonal perception in Western music and has influenced computational approaches in music theory and cognitive science.

\subsection{Brain-inspired Models} 
Facing the gap between technical proficiency and true musical comprehension, brain-inspired methods integrate brain mechanisms with the strengths of traditional AI models to enhance the model's cognitive abilities. Some researchers~\cite{Brain2Music} have utilized fMRI responses of human brain activity to music stimuli as embeddings for deep learning models. However, the primary goal of such studies has been to reconstruct music, rather than to explore the deeper aspects of musical understanding or creative generation. Spiking Neural Networks (SNNs), whose neurons and learning principles more closely resemble those of the human brain, offer a potential solution to the current challenges. Q. L. and colleagues proposed a spiking neural network inspired by cortical structures to memorize music ~\cite{LQ2020}, and to generate stylistic ~\cite{LQ2021} and emotional ~\cite{BrainCog} musical melodies. However, these studies primarily focus on monophonic melody generation and lack the integration of music theory. Overall, the application of brain-inspired spiking neural networks for learning and generating music remains rare. 
\begin{table*}[!h]
    \centering
    \caption{The musical terms used in this work}
    \begin{tabular}{p{2.6cm}|p{1.4cm}|p{11.8cm}}
        \toprule
         Term &  Notation & Description\\
        \midrule
         Pitch & p &  The frequency of a sound.\\
         Duration & d & The time interval of sound lasts.\\
         Note & N & A specific musical sound with a particular pitch and duration.\\
         Scale & Sca &  A series of individual notes proceed stepwise in ascending or descending order \\
         Mode & M & A scale arranged according to specific interval relationships. \\
         Key & K & The group of pitches, or scale, that forms the basis of a musical composition, and indicates the tonal center of the piece. \\
         Part & P & An individual line of music performed by a specific instrument or voice(or a group of instruments or voices).\\
         Chord & Ch & A combination of three or more notes played or heard together.\\
         Triads & Tr & Three-note chords made up of a root, third, and fifth.\\
         Diatonic Tone & DT & A note belongs to the specific seven-note scale of a major or minor key.\\
         Chromatic Tone & CT & A note that does not naturally belong to the seven-note diatonic scale of a given key.\\
         
         \bottomrule
    \end{tabular}
    \label{tab:terms}
\end{table*}

\section{Methodology}\label{sec:3}
\subsection{Model Architecture}
\begin{figure*}[!h]
  \centering
  \includegraphics[scale = 0.38]{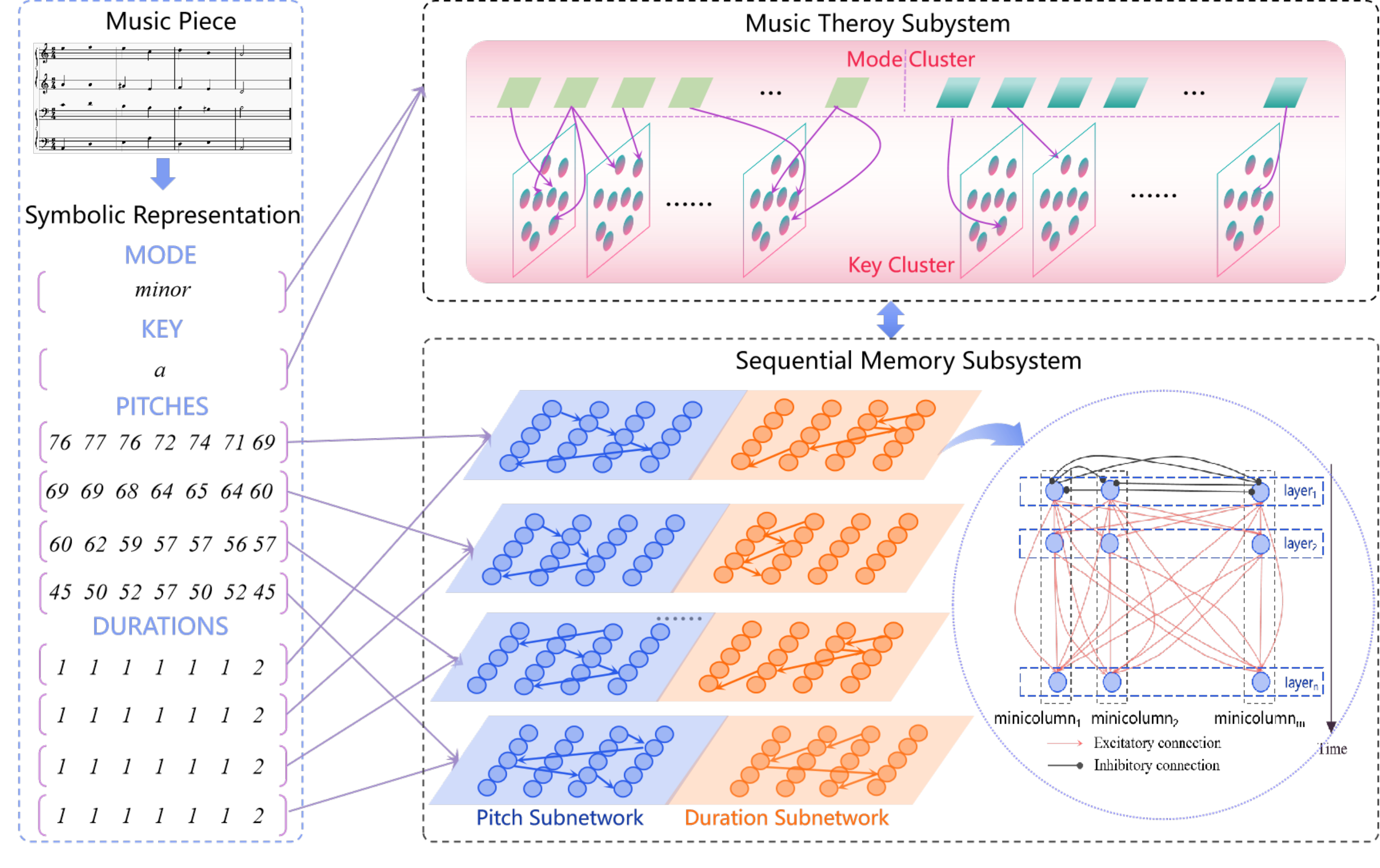}
  \caption{The architecture of the brain-inspired mode learning spiking neural network. The model contains the music theory subsystem(MTS) and the sequential memory subsystem(SMS). The MTS contains the mode cluster and key clusters, which is responsible for encoding the related music knowledge. The SMS receives the symbolic representation of the pitches and the durations, encoding and memorizing the relationships of the ordered notes. The structure of the pitch and the duration subnewtorks are marked by the dashed circles. }
  \label{fig:architecture}
\end{figure*}

In this study, we employ symbolic representations of musical pieces as the dataset and introduce a brain-inspired model based on a spiking neural network that is inspired by recent advancements in neuroscience and psychology in the music domain.  As illustrated in Figure \ref{fig:architecture}, the music information is prepared in a symbolic manner and divided into two parts: 1) treating the mode and key of a musical piece as theoretical musical knowledge, and 2) describing ordered notes, along with their respective pitches and durations with MIDI standard. Then, the model centers around two key components: a musical theory subsystem that embeds modes and chords as foundational prior knowledge directing the subsequent learning task, and a sequential memory subsystem designed to learn and store the ordered notes of music pieces.

\subsubsection{Music Theory Subsystem}
\begin{figure}
  \centering
  \includegraphics[width=1.0\columnwidth]{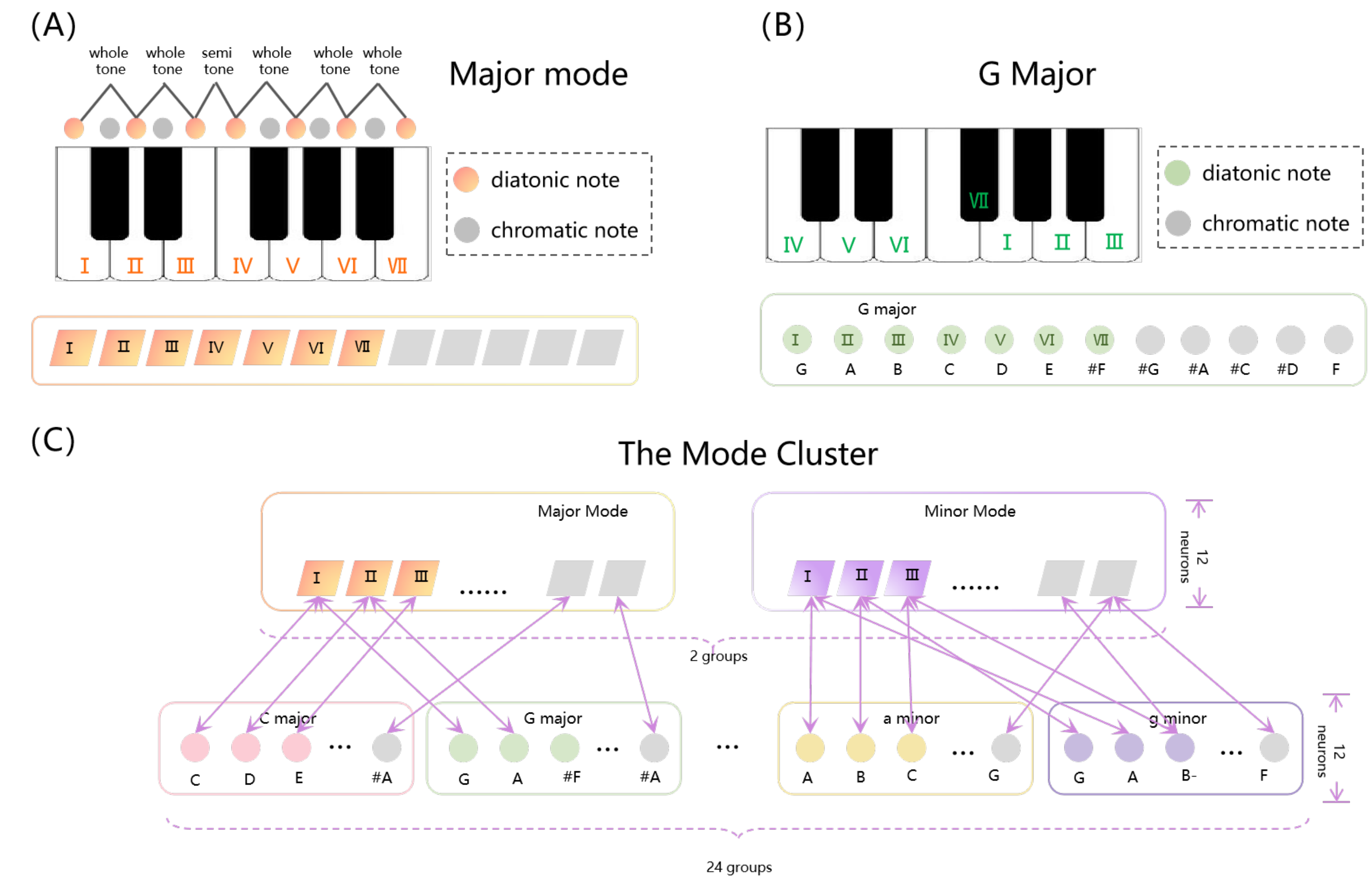}
  \caption{The musical theory subsystem, \textbf{(A)} describes how a neuron group in first layer in the mode cluster represents the major mode. Seven neurons drawn by orange parallelograms encodes the diatonic tones and the gray ones encodes the chromatic tones; \textbf{(B)} employs the key of G major to illustrate the principle, neurons drawn by green parallelograms encodes diatonic tones, G, A, B, C, D, E, \#F, the gray parallelograms also encodes the rest tones; \textbf{(C)} draws the hierarchical connection architecture of the mode cluster.}
  \label{fig:mode}
\end{figure}

The music theory subsystem(MTS) is structured hierarchically to encode modes as musical prior knowledge. As shown in Figure.~\ref{fig:mode}D, the first layer, the mode cluster, contains two neural groups dedicated to encoding the Western major and the minor modes. Each group in this layer consists of twelve neurons corresponding to the tones in 12-Tone Equal Temperament (12-TET), with seven of them representing the tones (\uppercase\expandafter{\romannumeral1} to \uppercase\expandafter{\romannumeral7}) within the pattern and the rest ones encoding those outside of the pattern, figure~\ref{fig:mode}A illustrates the details of the major mode representation. The second layer called the key cluster is consisted of twenty-four neural groups encoding twelve major keys and twelve minor keys. Similarly, each group comprises twelve notes with a specific tonic, figure~\ref{fig:mode}B shows how the neural group encodes the G major. Synaptic connections(shown in figure~\ref{fig:mode}C) are projected from neurons in each group in the first layer that represent the tone scale degree of the mode to those in the second layer encoding the corresponding note scale degree.\\

\subsubsection{Sequential Memory Subsystem}
A musical piece is a flow of ordered notes arranged across multiple instrumental parts or tracks, every track plays their harmonic role and contributes to the overarching harmony. For maintaining their individual characteristics, the sequential memory subsystem(SMS) is partitioned into four segments to encode and store the sequential order and temporal relationships of the notes, including melodic lines and harmony progressions. As depicted in figure.~\ref{fig:architecture}, each segment resides a dual-network configuration: a pitch subnetwork encoding the tones of notes, and a duration subnetwork retains the time intervals indicating how long a tone sustains. 

\paragraph{\textbf{The Pitch Subnetwork}}
Inspired by the neural populations in the primary auditory cortex that respond distinctively to different frequencies, the pitch subnetwork is proposed to consist of $128$ functional minicolumns as its building blocks. As depicted in Figure~\ref{fig:architecture}, each minicolumn is configured as a vertical column, representing one of the $128$ pitches in accordance with the MIDI (Musical Instrument Digital Interface) standard. Each minicolumn comprises numberous neurons, all of which share a common preference for a specific MIDI pitch index. For instance, all neurons within the minicolumn indexed at $60$ will respond preferentially to the pitch C4. The synaptic connections within and between the minicolumns are detailed below.

\paragraph{\textbf{The Duration subnetwork}}
The duration subnetwork is structurally identical to the pitch subnetwork. It consists of 64 functional minicolumns, each representing a particular note duration. These minicolumns are organized to cover a range of time intervals, with the finest granularity being a sixty-fourth note.  

\paragraph{\textbf{Intra-Connection}} \label{sec:intraconnection}
As illustrated in Figure~\ref{fig:architecture}, the pitch and duration subnetworks share the same internal connection structure. The synaptic connections between adjacent layers and across layers are excitatory and fully connected. Meanwhile, the introduction of synaptic plasticity and transmission delay plays a crucial role in the learning process. Synaptic plasticity allows the connection weights between neurons to be adjusted based on experience and learning, mirroring the adaptive capabilities of biological neural networks. Transmission delay, which accounts for the time it takes for signals to travel between neurons, is also incorporated to ensure that the timing of neural firing patterns is accurately represented. Connections within the same layer are inhibitory. For further details about the sequential memory system, see previous works~\cite{LQ2020,LQ2021}.

\subsection{Music Data Encoding}
The encoding process aims to transform external stimuli to neural spikes inspired by brain mechanisms. Suppose a music piece consisting of multiple parts is defined as $NS=\{N_{i,j}|i=1,2,...n_{j},j=1,2,3,4\}$, where $N_{i,j}$ denotes the note at $i$-th position in the $j$-th part, $n_{j}$ refers to the number of notes in the $j$-th part. The mode can be written as $M_{r}$, where $r=0$ or $r=1$ refers to the major or minor mode. The key is marked as $K_{s}$, where $s\in[1,12]$ denotes one of the 12 possible tonal center. Then, the encoding process can be divided into two steps:
\paragraph{\textbf{Step1:Transformation of External Stimuli}}
The external stimuli(such as keys, pitches,etc.) are transformed into the input current $I$ that is suitable for the computational neuron model to receive. For example, a G-Major musical piece activates the neurons in the major cluster in the first layer and G major in the second layer. Equation.\ref{eq1}-.\ref{eq2} describes the transformation for the mode and key clusters, respectively. 
\begin{align}
    \label{eq1}
    I^{M_{r}}_{E\_i}(t) &= \alpha^{M_{r}}\delta(x_{ij}(t)-sd_{i}^{M_{r}})\\
    \label{eq2}
    I^{K_{s}}_{E\_i}(t) &= \alpha^{K_{s}}\delta(x_{ij}(t)-ts_{i}^{K_{s}})
\end{align}
where $I_{E\_i}^{M_{r}}(t)$ is the input current for neuron $i$ in the $r$-th group of the mode cluster at time step $t$. $x_{ij}(t)$ represents the external stimuli, which refers to the pitch of the input note $N_{i,j}$ at time $t$. $sd_{i}^{M_{r}}$ denotes the scale degree of the tone which the neuron $i$ represents in the mode cluster. Similarly, the $I^{K_{s}}_{E\_i}(t)$ refers to the input current of neuron $i$ in the $s$-th group of the key cluster, $ts_{i}^{K_{s}}$ denotes the tone scale in G major cluster which neuron $i$ represents.$\alpha^{M_{r}}$= $\alpha^{K_{s}}$= $50$ are the scale factors to control the input values.\\
The sequential memory subsystem is responsible for transforming the pitch and duration of ordered notes by the pitch and duration subnetworks in the $j$-th part, the current are as follows:
\begin{align}
    \label{eq3}
    I^{P}_{E\_ij}(t) &= \alpha^{P}\delta(x_{ij}(t)-p_{hj})\\
    \label{eq4}
    I^{D}_{E\_ij}(t) &= \alpha^{D}\delta(y_{ij}(t)-d_{kj})
\end{align}
where $x_{ij}(t)$ and $y_{ij}(t)$ are the pitch and duration of the note $N_{i,j}$ at time step $t$, $p_{hj}$ and $d_{kj}$ denotes the preferences of the neurons in the $h$-th and $k$-th minicolumns in the pitch and duration subnetworks of part $j$, respectively. $\alpha^{P}$= $\alpha^{D}$ = $30$ controls the scale of the current.  
\paragraph{\textbf{Step2:Neural Spiking}}
In this study, we use the Izhikevich neural model~\cite{E2003Simple} to simulate the behaviors of neurons within both the musical theory subsystem and the sequential memory system. The model is described as follow equations:\\
\begin{equation}
    \begin{aligned}
        v(t+1) &= \begin{cases}
        0.04v(t)^{2} + 5v(t) + 140-u(t) + I(t), v(t) < V_{th}\\
        c, v(t) \geq V_{th}
        \end{cases}\\
        u(t+1) &= \begin{cases}
        a(bv(t)-u(t)), v(t) < V_{th}\\
        u(t)+d, v(t) \geq V_{th}
        \end{cases}\\
    S(t) &=  \begin{cases}
     0, v(t) < V_{th} \\
     1, v(t) \geq V_{th}
    \end{cases}
    \end{aligned}
\label{eq5}
\end{equation}
$v(t+1)$ and $v(t)$ represent the membrane potential at time $t+1$ and $t$, $u(t+1)$ and $u(t)$ denote the recovery variable at time $t+1$ and $t$, respectively. $I(t)$ is the synaptic current input to the neuron at $t$. $a$, $b$, $c$, and $d$ are parameters that control the model to fire with different spiking patterns. When the membrane potential reaches the threshold $V_{th}$, it is reset to $c$ and the recovery variable $u(t)$ is incremented by $d$. $S(t)$ represents whether the neuron exhibits a spike at time $t$. 
Neurons in the music theory and sequential memory subsystems are simulated by Izhikevich neural model in response to the input current $I$ and deliver the spikes. In this study, we set the parameters $a=0.1$, $b=0.2$, $c=-65$, $d=30$, $V_{th} = 30$.

\subsection{Learning Based On Neural Circuits Evolution} 
The brain utilizes multiple regions to work together, with dynamic synaptic formation and elimination creating diverse neural circuits to accomplish various cognitive tasks. Inspired by these mechanisms , we consider the learning process as a collaborative effort among interconnected subnetworks, with a focus on the dynamic neural circuit evolution over time. 
\subsubsection{Synaptic Creation}
 Since the music theory system stores the prior knowledge, the inter-connected architecture is preset. However, there are no synaptic projections between the musical theory subsystem and the sequential memory system at the initial state. 

 Neuroscientific research has demonstrated  that neural electrical activities and the dynamics of axonal growth cone movements are interdependent and coordinated. Electrical activities serve as feedback signals for the navigation of growth cones, while the movement of growth cones and the formation of new synapses subsequently influence the electrical activities within the neural network.~\cite{spitzer2006electrical, HuaSmith2004, ZhengPoo2007}. In this study, we simplify this complex process by establishing rules for the formation of new synaptic connections and neural circuits in response to the input of musical information.
\begin{equation}
    o = \sum_{f}^{N}\sum_{n}^{N}\delta(t_{i}^{f}-t_{j}^{n})
    \label{eq6}
\end{equation}
\begin{equation}
    c_{ij} = \begin{cases}
        1, o \geq 5\\
        0,else
    \end{cases}
\label{eq7}
\end{equation}
where, the $t_{i}^{f}$ and $t_{j}^{n}$ represent the spike time $f$ and $n$ of postsynaptic neuron $i$ and presynaptic neuron $j$, respectively.  When the oscillatory times $o \geq 5$ of these neuron $i$ and $j$, a new synaptic connection is formed, and is represented as $c_{ij}$. This simple rule is defined to describe that neurons oscillate together are more likely to form a synaptic connection between them.
\begin{figure}
  \centering
  \includegraphics[width=1.0\columnwidth]{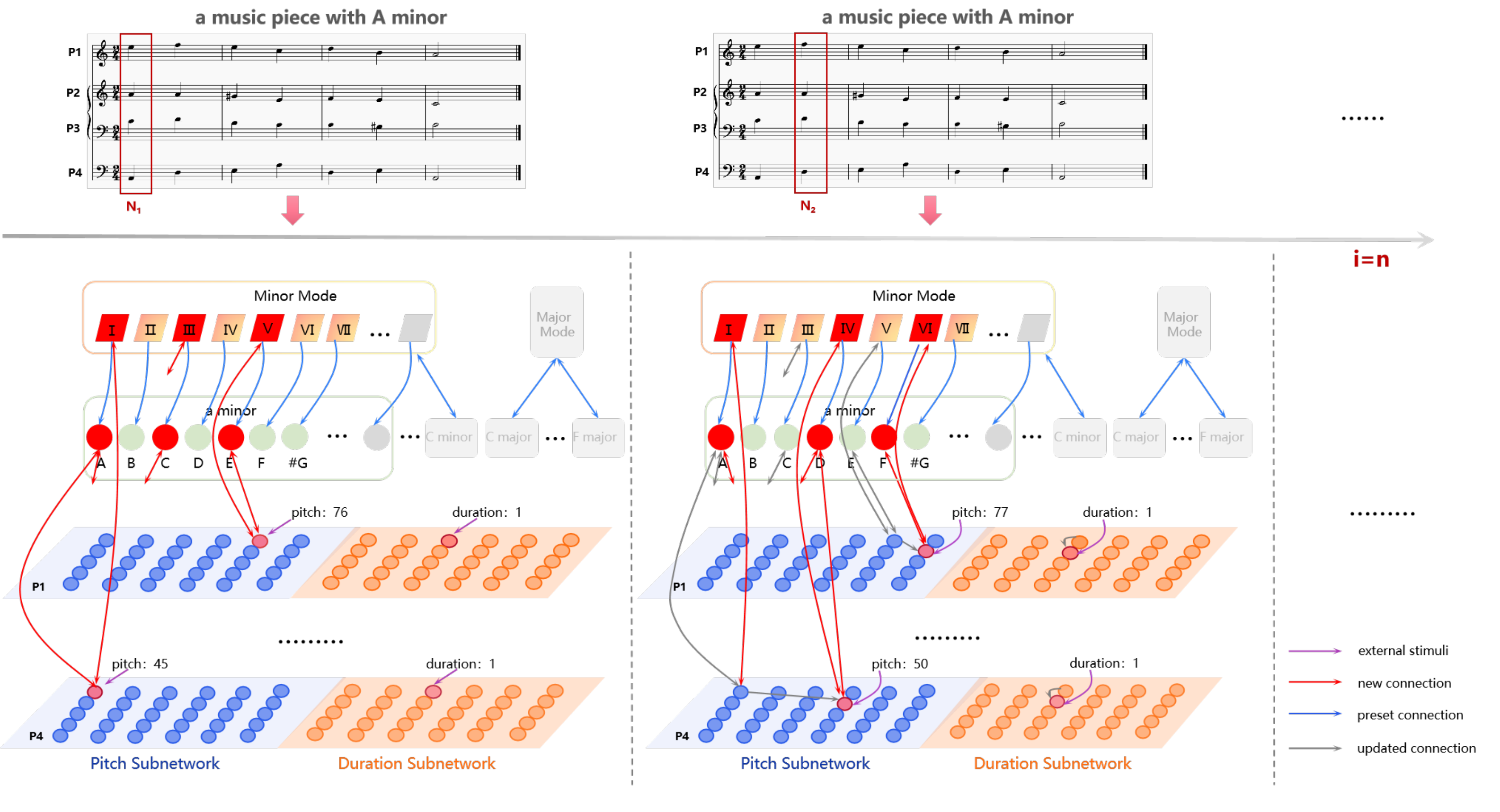}
  \caption{The learning process of music guided by the mode theory.}
  \label{fig:learning}
\end{figure}

The establishment of novel neural pathways mainly occurs between the musical theory subsystem and the sequential memory subsystem, which indicates that the collaborative learning and interaction of these two subsystems. Figure.~\ref{fig:learning}  illustrates a simple example of how the model learns a four-part music piece in A minor. The model starts by receiving a set of notes $N_{1}={\{N_{1,1},N_{1,2},N_{1,3},N_{1,4}\}}$ from four parts, accompanied by their respective symbolic pitches $X_{1}= \{{76,69,60,45\}}$. These notes fire the corresponding neurons in the mode and key clusters (highlighted by red parallelograms and circles) according to Eq.\ref{eq1}-\ref{eq2} and Eq.\ref{eq5}. Simultaneously, the notes $N_{1}$ also prompt the activation of neurons(represented by red circles) situated across various minicolumns within the pitch subnetworks of the sequential memory subsystem. These neural oscillations facilitate the formation of both feedforward and feedback connections(marked as red double arrows) between neurons in the musical theory subsystem and those in the sequential memory system by the rule~\ref{eq6}. The synaptic weights are initially set at random values, reflecting the stochastic nature of early learning stages. These weights will be subsequently updated by STDP learning rule as the learning process unfolds. The transmission delay $del$ for these new connections is set to zero in this phase, implying instantaneous signaling for simplicity. It is worth noting that the prior knowledge relationships, embodied as synaptic connections, are already established and depicted by blue arrows. Upon receiving the subsequent note set $N_{2}$, a similar pattern ensues, triggering the activation of corresponding neurons across both subsystems. When no connections link two oscillating neurons, novel connections are established spontaneously. This intricate process is characterized by the creation of neural circuits spanning the two subsystems.

\subsubsection{Synaptic Plasticity}
This paper utilize the modified STDP(Spike-Timing Dependence Plasticity)\cite{bi1998synaptic} learning rule to account for the spiking transmission delay when updating synaptic weights during the learning process, as shown in Eq.~\ref{eq8}. 
\begin{equation}
	\begin{split}
		&\Delta w_{j}=\sum\limits_{f=1}^{N} \sum\limits_{n=1}^{N} W(t^{f}_{i}-t^{n}_{j}-t^{del}_{i,j})\\
		&W(\Delta t)= 
		\left\{
		\begin{split}
		A^{+}e^{\frac{-\Delta t}{\tau_{+}}}\quad if\; \Delta t>0&\\ 
		-A^{-}e^{\frac{\Delta t}{\tau_{-}}}\quad if\;\Delta t<0&\\ 
		\end{split}
		\right.
	\end{split}
	\label{eq8}
\end{equation}
where $\Delta w_{j}$ is modulate weight of the synapse $j$. $t^{f}_{i}$ and $t^{n}_{j}$ denotes the spike time of post-synaptic neuron $i$ and pre-synaptic neuron $j$, respectively. $t^{del}_{i,j}$ represents the axonal transmission delay between this neuron pair. $A_{+}$, $A_{-}$ are scale factors, and $\tau_{+}$, $\tau_{-}$ are time constants.

As depicted in Fig.~\ref{fig:learning}, following the emission of spikes by neurons within the pitch and duration subnetworks triggered by the input of the initial note set $N_{1}$, which comprises a pitch set $X_{1} = {\{76, 69, 60, 45\}}$ and a duration set $Y_{1} = {\{1, 1, 1, 1\}}$, the model proceeds to process the subsequent note set $N_{2}$. Upon arrival of $N_{2}$, characterized by its pitch set $X_{2} = {\{77, 69, 62, 50\}}$ and identical duration set $Y_{2} = {\{1, 1, 1, 1\}}$, the corresponding neurons generate spikes in distinct minicolumns, each with its own preference for different parts of the input (highlighted by red circles).It becomes evident that neurons in the sequential memory system receive inputs not just from the external stimuli, but also from other layers within the same subnetworks and from the theory subsystem, integrating this information for further processing. Then, the inputs of the $i$-th neuron in the $j$-th minicolumn in the pitch subnetwork can be described as Eq.~\ref{eq9}.
\begin{equation}
    I_{ij}^{P}(t+1) = I^{P}_{E\_ij}(t)+\sum_{s}w^{MTS}_{si}(t)+\sum_{r}w^{P}_{ri}(t)
    \label{eq9}
\end{equation}
$I_{ij}^{P}(t)$ denotes the total input of the $i$-th neuron in $j$-th minicolumn located in the pitch subnetwork at time step $t$. $I^{P}_{E\_ij(t)}$ is the current caused by the external stimuli(pitch value), $w^{MTS}_{s}$ means the $s$-th synaptic weight from the music theory subsystem, and $w^{P}_{ri}$ represents the $r$-th synaptic weight from the neurons in pitch subnetwork. The detailed discussion regarding this compounent can be found in our earlier work~\cite{LQ2020}. Similarly, the input of neurons in the duration subnetwork can be expressed by Eq.~\ref{eq10}. A notable difference, however, is that neurons in duration subnetwork do not receive the signals from music theory subsystems.
\begin{equation}
    I_{ij}^{D}(t+1) = I^{D}_{E\_ij}(t) +\sum_{r}w^{D}_{ri}(t)
    \label{eq10}
\end{equation}

\subsection{Mode-Conditioned Music Generation}
\begin{figure}
  \centering
  \includegraphics[width=1.0\columnwidth]{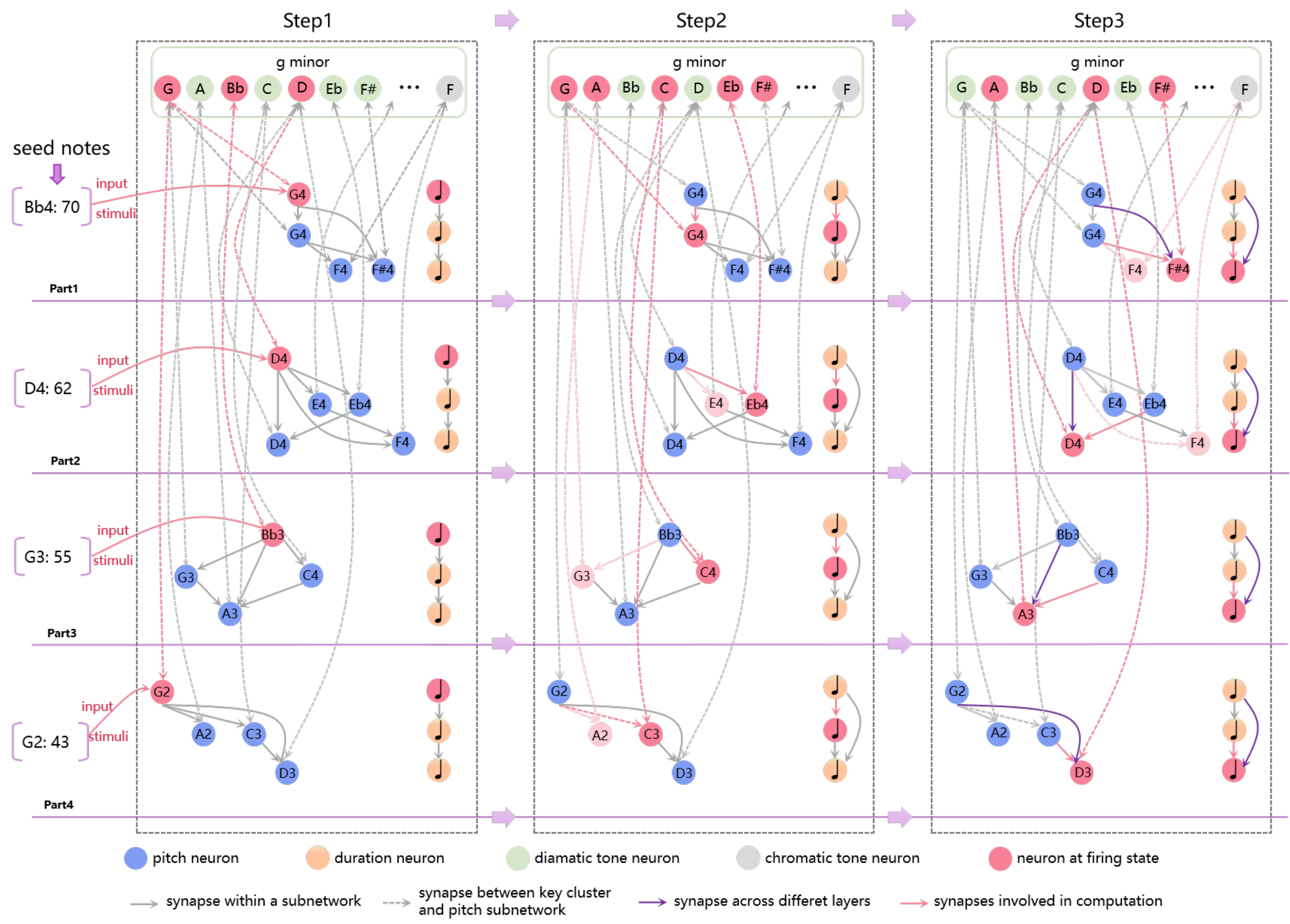}
  \caption{The generation process of a musical piece in G minor begins with a tonic chord as the seed. The model receives this seed input and activates the neurons representing pitches that distribute in four parts, guided by the active neurons that represent G minor, step by step. Neurons and synaptic connections not involved in the generation are omitted for clarity. Green circles represent neurons within the G minor group of the key cluster, blue and orange circles denote the pitch and duration neurons in the sequential memory subsystem, and red circles mark neurons activated at different time steps.}
  \label{fig:generation}
\end{figure}

Specifying the mode and key is a fundamental step in the process of composing a piece of music. The model described in this paper requires not only the mode and key but also a set of seed notes to initiate the creative process. As illustrated in Fig.~\ref{fig:generation}, the model is tasked with generating a four-part musical piece in G minor, starting with the tonic chord(G2-Bb3-D4-G4) as the seed. The figure omits the neurons and synaptic connections that are not involved in this example. The tonic chord is defined by the notation $N_{0}$, which includes the pitch set $X_{0}={\{67,62,58,43\}}$, and duration set $Y_{0}={\{1,1,1,1\}}$, sending to the corresponding parts of the pitch and duration subnetworks.  
\begin{itemize}
    \item {\textbf{Step1:}}
    Upon with the input of the initiate seed notes, the corresponding neurons in the G-minor group of key cluster are ready to receive the stimuli and emit the spikes(red circle). The pitch and duration values of the seed notes trigger the activation of specific neurons in sequential memory system(blue and orange circles), which then propagate through the network to generate the next set of notes. At this step, the neurons are updated by Eq.~\ref{eq2}-Eq.~\ref{eq5}. Neurons representing G2, Bb3, D4 and G4 in part one to part four fire(marked by red circles) and propagate their spikes to other neurons. 
    \item {\textbf{Step2:}}
    Then, the subsequent neurons of the pitch and duration subnetworks receive and integrate inputs through trained synaptic connections by Eq.\ref{eq11} and Eq.\ref{eq12}, respectively.
    \begin{equation}
        I^{P}_{ij}(t+1) = \sum_{s}w^{MTS}_{s}(t) + \sum_{r}w^{P}_{r}(t)
        \label{eq11}
    \end{equation}
    \begin{equation}
        I_{ij}^{D}(t+1) = \sum_{r}w^{D}_{ri}(t)
        \label{eq12}
    \end{equation}
    Therefor, neurons in the same layer compete with each other and we employ the Winner-Takes-All priciple to select the most strongly activated neuron as the generated results. As illustrated in Fig.~\ref{fig:generation}, neurons in part two representing E4 and Eb4 are both activated as candidates at this step. However, due to the connection architecture of our trained model between the key cluster and pitch subnetworks, which is similar to KS model, and because the tone E is a chromatic tone in G minor, the synaptic weight between the neuron representing E4 and the neuron representing tone E in G minor group is significantly lower(light pink arrow) compared to that between neuron Eb4 and the corresponding neuron Eb in the G minor group of the key cluster(red arrow). Consequently, the neuron Eb4 emits more spikes and becomes the winner, representing the next generated note in part two. The generating process are similar in other parts, synapses with high weights are marked by red arrows. C4 and C3 are the final winner in part three and part four. In summary, the neuron with the highest firing rate as the winner in each layer of each part is described by Eq.~\ref{eq13}.
    \begin{equation}
        \label{eq13}
        z = \mathop{arg\max}_{j}(\sum^{T}_{t}S_{j}(t))
    \end{equation}
    Here, $z$ represents the index of the winning neuron, $S_{j}(t)$ denotes the spike train of neuron $j$ at time $t$, and $T$ is the total duration over which spikes are counted.
    Overall, in this step, the generated notes are G4, Eb4, C4, C3 in respective part.
    \item {\textbf{Step3:}}
    The model generates consequent notes by the similar computation process to step two. The difference at this step is that, connections across different layers are involved in pitch and duration subnetworks, which are marked by the purple arrows. The final results in this step from part one to part four are F\#4, D4, A3 and D3, respectively.
\end{itemize}

\section{Datasets}
\subsection{Dataset Description}\label{sec:datasets}
Human learners typically begin studying music theory with a textbook, starting with fundamental concepts and completing exercises to build foundational knowledge. Through repeated practice and exposure, they solidify this knowledge by listening to numerous musical works, which helps reinforce their memory. With this prior knowledge in place, they can progress to analyzing compositions or even creating their own music. Inspired by human behavior, this paper utilizes two datasets to train our model:

\textbf{Sposobin's Harmony Textbook Exercises (SHTE)}:
To facilitate an in-depth and direct study of modal and tonal features, this paper introduces a new dataset comprising 193 four-part harmony excerpts, carefully selected and annotated from Chapters 1 to 19 of \emph{Sposobin's Harmony Textbook}. Each excerpt represents an accurate solution crafted by music professionals and is stored in XML format to support efficient digital processing. The strength of this dataset lies in its conciseness and richness in music theory, with each excerpt designed to highlight a specific point. It includes 96 music pieces in major keys and 97 in minor keys, with the absence of Eb and Ab minor. The number of music pieces in each key of this dataset are detailed in Table.\ref{tab:keydata}

\textbf{J.S. Bach's four-part chorales(Bach)}:
To make the model 'listen' more music works, we employ the J.S. Bach's four-part chorales dataset, which is available through the Music21 python package~\cite{cuthbert2010music21}. It includes 408 famous chorales in \emph{musicxml} format, with 219 in major keys and 189 in minor keys, excluding Db and Gb major, as well as Db, Eb, and Ab minor. A summary of the key distributions is provided in Table.\ref{tab:keydata}

Both datasets provide a valuable resource of material for training and evaluating our music generation model, particularly focused on generating four-part harmony in various modes and keys. 
\begin{table*}[!h]
    \centering
    \caption{The number of keys in SHTE and Bach datasets}
    \begin{tabular}{c|c|c|c|c|c|c|c|c|c|c|c|c|c}
         \hline
         \hline
          \multicolumn{2}{c|}{Keys}&C & C\#/Db & D & D\#/Eb & E & F & F\#/Gb & G & G\#/Ab& A & A\#/Bb & B \\
         \hline
         \multirow{2}{*}{Major}& SHTE & 6 &4 & 11 & 8 & 10 & 13 & 1 & 8 &8 & 11 & 13 & 3\\
         \cline{2-14}
         \multirow{2}{*}{} &Bach & 22 & - & 32 & 9 & 10 & 28 & - &59 & 1 & 32 & 25 & 1\\
         \hline
         \hline
         \multirow{2}{*}{Minor}& SHTE & 9 &6 & 10 & - & 11 & 7 & 17 & 17 &- & 9 & 2 & 9\\
         \cline{2-14}
         \multirow{2}{*}{} &Bach & 10 & - & 27 & - & 20 & 2 & 7 &44 & - & 49 & 2 & 28\\
         \hline
         \hline
    \end{tabular}
    \label{tab:keydata}
\end{table*}

\section{Experiments}
\subsection{Our Model VS Psychological Model}
As described in Section \ref{sec:2}, the Krumhansl-Schmuckler model(KS model) highlights the significance of pitch classes within each major and minor key. In our model, the synaptic architecture between mode cluster and the sequential memory system, particularly the pitch subnetwork,  mirrors the importance of neural relationships, indicating the critical role of each pitch class within a key. To observe the internal structure of our model, this paper proposes two features which are listed below: 
\begin{itemize}
    \item {\textbf{Pitch Synaptic Count(PSC)}}
     refers to the total number for each neuron representing different pitch class in the mode clusters and the corresponding neurons in the pitch subnetwork of the sequential memory system. Since the $k$-th neuron reprsents the $k$-th pitch class, the PSC for the the $k$-th pitch lass can be computed as Eq.~\ref{eq14}.
    \begin{equation}
    PSC_{k} = \sum^{4}_{j=1}\sum^{128}_{r\%12=k}\sum^{N}_{s}c_{k,jrs}
    \label{eq14}
    \end{equation}
    where $c_{k,jrs}$ denotes whether a synapse exists between the $k$-th neuron in each neural group of the mode cluster and the $r$-th neuron in the $s$-th minicolumn of the $j$-th part in the pitch subnetwork(see Eq.\ref{eq7}). $PSC(k)$ represents the pitch synaptic count between the $k$-th neuron, corresponding to the related pitch class in the major and the minor modes, and all the neurons belonging to the same pitch class in the pitch subnetwork.
    \item {\textbf{Pitch Average Synaptic Weights(PASW)}}
    calculates the average weights of synapses between each pitch neuron in the mode/key clusters and the pitch subnetwork.
    The computation is as Eq.~\ref{eq15}
    \begin{equation}
    PASW(k)= \frac{\sum^{4}_{j=1}\sum^{128}_{r\%12=k}\sum^{N}_{s}w_{j,r,s}(k)}{PSC(k)}
    \label{eq15}
    \end{equation}
     where, $w_{j,r,s}$ is the synaptic weight between the $k$-th neuron in each neural group of the mode cluster and the $r$-th neuron in the $s$-th minicolumn of the $j$-th part in the pitch subnetwork, $PASW(k)$ denotes the average synaptic weight between the $k$-th pitch class for each mode and related neurons in the pitch subnetwork, $PSC_{k}$ is computed as the Eq.~\ref{eq14}.
\end{itemize}

\begin{figure*}[!h]
  \centering
  \includegraphics[width=1.0\textwidth]{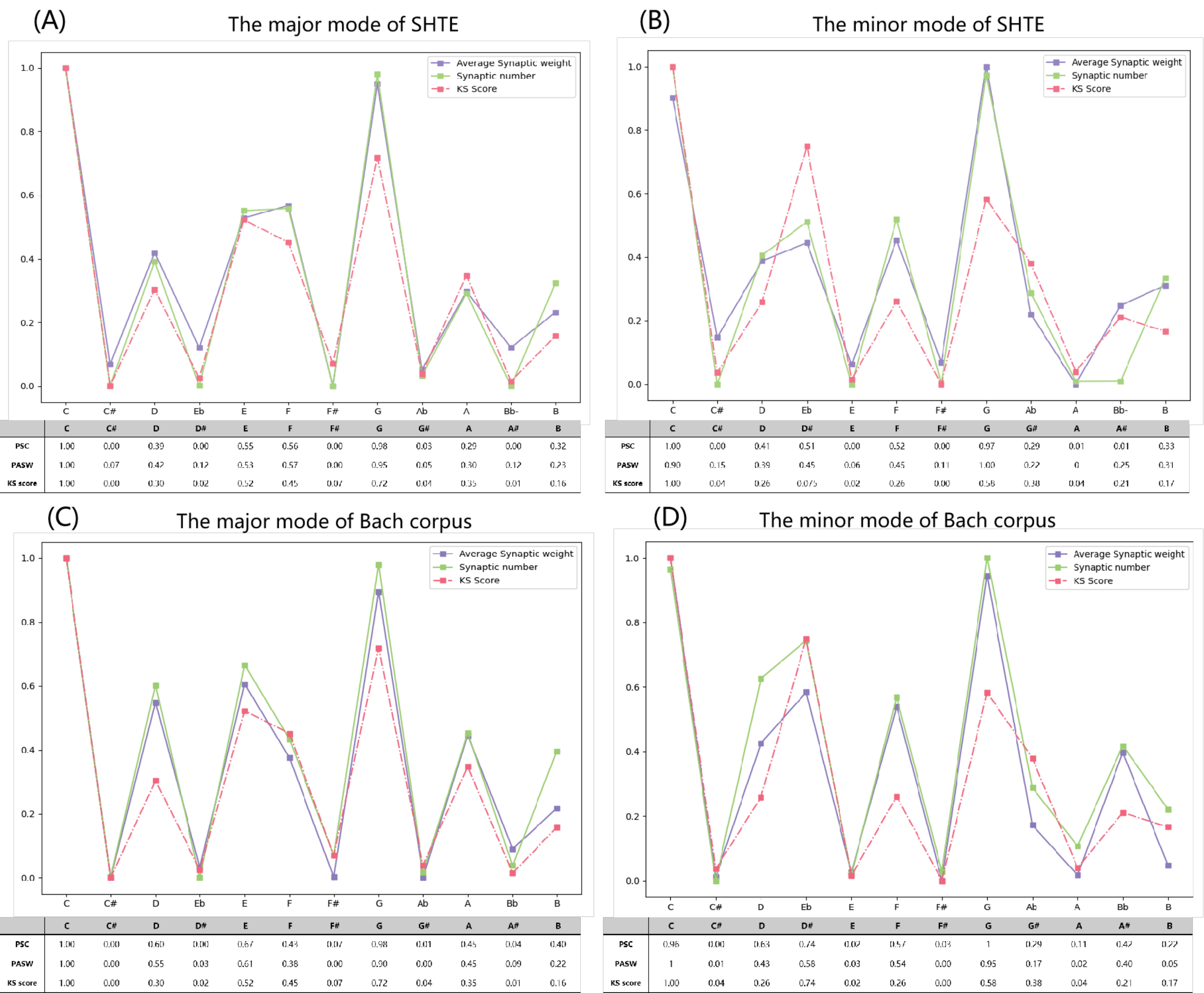}
  \caption{The comparative analysis between the connection architecture of our model and the Krumhansl-Schmuckler model is presented as follows: Panel (A) shows the average synaptic weight and synaptic count of neurons within the major cluster and pitch subnetwork trained on the SHTE dataset, while panel (B) illustrates these metrics for the minor cluster and the pitch subnetwork, also trained on SHTE. Additionally, panels (C) and (D) display the same metrics for both major and minor clusters and pitch subnetworks trained on the Bach corpus.}
  \label{fig:modeSyn}
\end{figure*}

To observe the changing trend in comparison with the KS model profile, we train our model using both SHTE and Bach datasets. After training, we normalize to PSC, PASW and KS pitch profile scores. Fig.\ref{fig:modeSyn} panels (A) and (B) present the quantified results for the major and minor modes trained on the SHTE, while panels (C) and (D) show results for training on the Bach corpus. In each picture, the purple and green lines represent the average synaptic weights and the total counts across the 12 pitch classes for the major and minor modes, respectively. While the pink dashed line draws the KS pitch profile scores. 

For the major mode in both dataset(Fig.\ref{fig:modeSyn}(A)(C)), the PSCs and PSWAs are consistent with the KS model, always receive the highest score $1.0$ at the point of tonic(C), followed by the dominant(G), with values $PSC_{SHTE}(G)=0.98$, $PASW_{SHTE}(G)=0.95$, and $PSC_{Bach}(G)=0.98$, $PASW_{Bach}(G)=0.9$, respectively. This pattern reflects the central roles of the tonic and dominant pitches. For the mediant tone, compared to the KS model $KS_{major}(E)=0.52$, the $PSC_{SHTE}(E)=0.55$,$PASW_{SHTE}(E)=0.53$, $PSC_{Bach}(E)=0.67$, $PASW_{Bach}(E)=0.61$, similar to the KS model. However, for the superdominant tones, the PSC and PSAW values in the Bach dateset are consistent with the KS model. In the SHTE dataset, however, these values are $PSC_{SHTE}(F)=0.56$ and $PASW_{SHTE}(F)=0.57$ at the subdominant(F), slightly higher than the values at the mediant tone(E), due to the frequent use of subdominant chords within our teaching dataset. The PSC and PSWA values for the remaining pitch classes also follow the same trend as observed in the KS model.

For the minor mode(Fig.\ref{fig:modeSyn}(B)(D)), the PSCs and PASWs also play the most important role at tonic(C). However, in the KS model, the scores for mediant(Eb)$KS_minor(Eb)=0.75$ are higher than those for the dominant(G)$KS_minor(G)=0.58$. In contrast, the PSC and PASW values in the SHTE dataset($PSC_{SHTE}(Eb)=0.52$, $PASW_{SHTE}(Eb)=0.45$) and Bach dataset ($PSC_{Bach}(Eb)=0.74$, $PASW_{Bach}(Eb)=0.58$) are lower than those at the dominant(G). This discrepancy can be attributed to the key and necessary role of the dominant chord in the cadences and its frequent use in both datasets. Besides, compared to KS model in SHTE dataset(\ref{fig:modeSyn}(B)), these two metrics for the leading tone(B) show an ascending trend,  This is due to the teaching characteristics of the SHTE dataset, which includes more exercises focused on learning chords.

This paper further calculates the cosine similarities between PSC and PASW with the KS scores, yielding values of $COS_{SHTE}(PSC_{major},KS_{major})=0.93$ and $COS_{SHTE}(PASW_{minor},KS_{minor})=0.92 $ for the SHTE dataset, while $COS_{Bach}(PSC_{minor},KS_{minor})=0.94 $ and $COS_{Bach}(PASW_{minor},KS_{minor})=0.94 $ for the Bach dataset, respectively. The results indicate a strong alignment between our model's connection architecture and the Krumhansl-Schmuckler psychological key perception model. The presence of subtle differences indicate that the model captures dataset-specific harmonic characteristics, suggesting that it not only aligns with KS-defined roles for tonic and dominant pitches but also adapts to dataset-specific harmonic nuances, such as the influence of subdominant or leading tones.

\begin{figure*}[!h]
  \centering
  \includegraphics[width=1.0\textwidth]{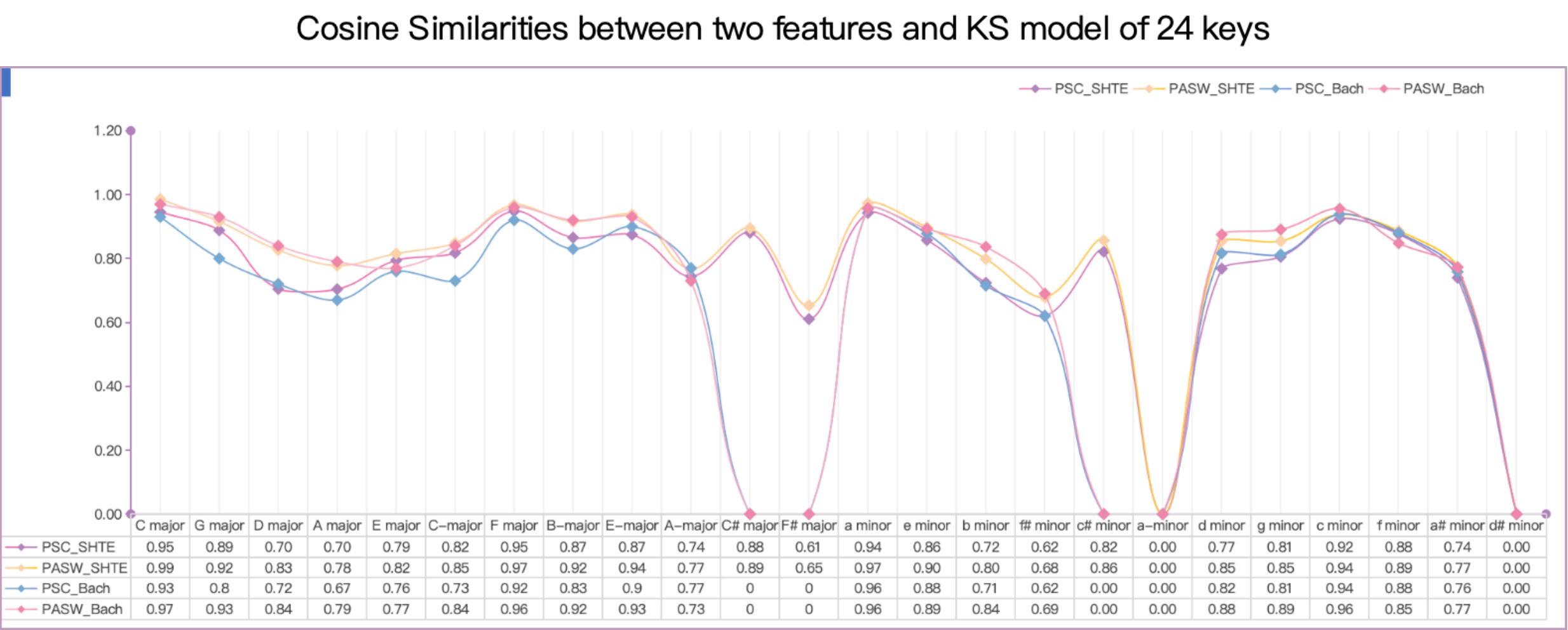}
  \caption{The results of cosine similarities for 24 keys between the two features(PSC and PASW) and the KS model under the SHTE and Bach Datasets.}
  \label{fig:24keys}
\end{figure*}

We also calculate the cosine similarities between these two metrics of the connection architecture and the KS model for the twelve major and minor keys. The result is provided in Fig.\ref{fig:24keys}.

This figure shows the cosine similarities between two features (PSC and PASW) derived from two datasets (SHTE and Bach) compared with the Krumhansl-Schmuckler (KS) model across 24 musical keys (12 major and 12 minor). The key points of analysis are:
\begin{itemize}
    \item The majority of values are above 0.7, indicating a strong alignment between the model's features (PSC and PASW) and the KS model.Specific keys, such as C major, G major, A minor, and C minor, exhibit particularly high cosine similarities across all datasets and features, affirming the model's robustness for these keys.
    \item For F\# major and F\# minor, the cosine similarities are consistently lower (below $0.7$), suggesting greater variability in the pitch content of musical pieces in these keys within the datasets. This phenorminan suggests a diversity in pitch usage or less structured tonality in the available pieces for these keys.
    \item Zero values are observed for C\# major, F\# major, C\# minor, Ab minor, A\# minor, and D\# minor. These results indicate that no musical pieces in these keys are present in the corresponding datasets (SHTE and Bach). This absence limits the evaluation of the model's performance for these keys.
    \item For the SHTE dataset, the PASW feature tends to have slightly higher similarities compared to PSC for most keys. While for the Bach dataset, both PSC and PASW demonstrate comparable trends, with slightly lower values for some keys compared to SHTE.    
\end{itemize}
Overall, the figure confirms that the developed model's features effectively capture key-related properties similar to human key perception (as modeled by the KS framework). Approximately $91.5\%$ of the values are above 0.7, showcasing the alignment's reliability.

\subsection{Generation Evaluation And Analysis}
\begin{figure}[!h]
  \centering
  \includegraphics[width=1.0\columnwidth]{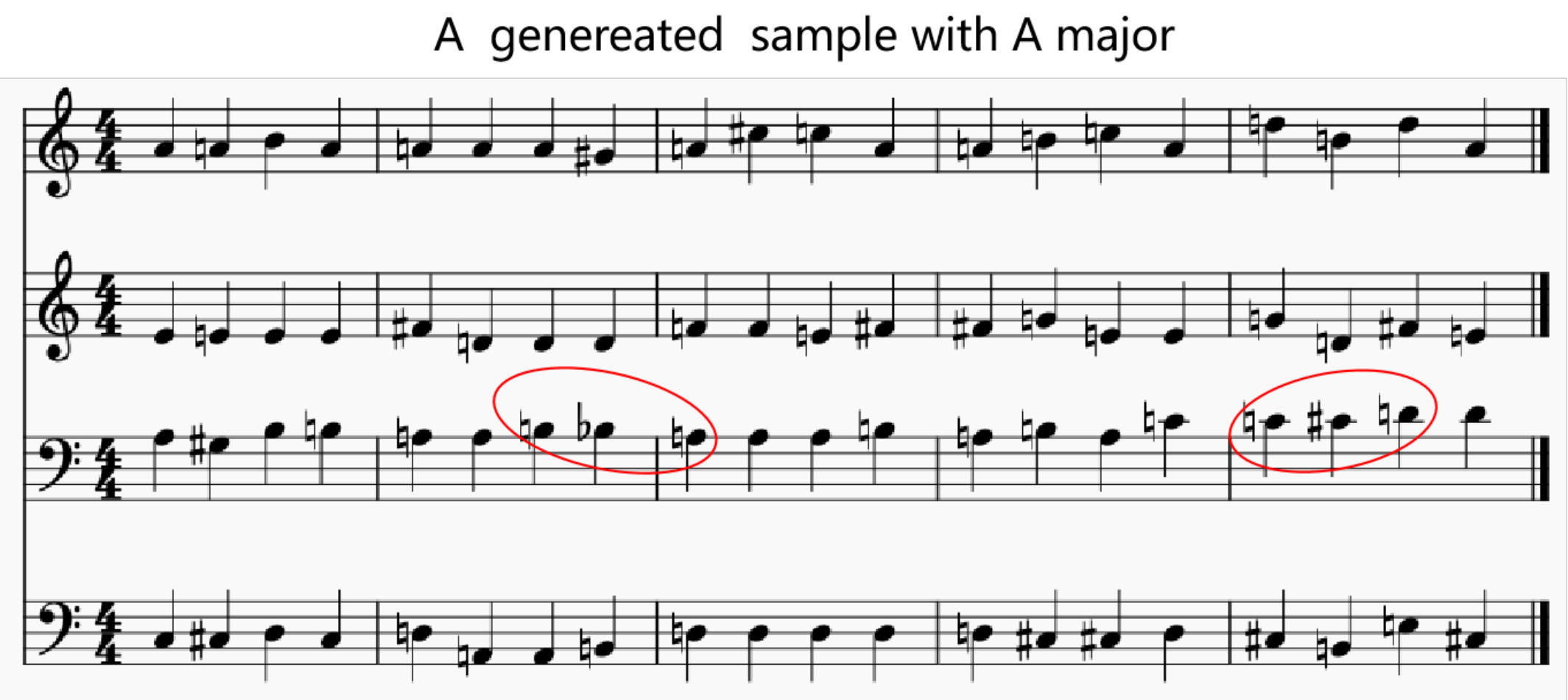}
  \caption{A generated sample in A Major, with red circles indicating semitone step ascending or descending for Smoother Melodic Transitions.}
  \label{fig:samples}
\end{figure}

In this section, the model generates musical pieces by specifying seed notes and setting the music length, with conditioning applied to various major and minor keys. Fig.\ref{fig:samples} presents an example of the generated samples from our model. The sample demonstrates that the generated tones of each part are basically accordance with the specified tonality. An intriguing phenomenon, marked by red circles, is that the model naturally learns to incorporate ascending or descending semitone steps, creating smoother melodic transitions. The use of semitones introduces chromatic tones, which break up the conditioning of key and the mode, enhancing expressiveness and allowing the generated melodies to explore more intricate harmonic nuances.

\subsubsection{Metrics}

To further assess and analyze the quality of the generated music pieces, we employ a comprehensive objective method which is inspired by the method proposed by Yang and Lerch~\cite{yang2020evaluation}. Several features are extracted from the generated samples and two datasets involved in our model. The features include: \textbf{Pitch Count(PC)}, the number of different pitches in a music piece; \textbf{Pitch Class Histogram(PCH)}, the histogram of 12 pitch classes; \textbf{Diatonic Pitch Rate(DPR)}, the proportion of pitches in a music piece that belongs to the key's diatonic scale;  \textbf{Pitch Range(PR)}, the pitch range between the highest and the lowest pitch in semitones within a sample;  \textbf{Pitch Interval(PI)}, average interval between two pitch neighbors, \textbf{PCTM}, the pitch class transition matrix.  

\subsubsection{Performance Comparison}
\textbf{A) Comparison with datasets}

Table.~\ref{tab:absmetrics} shows the comparison results between the generated samples and the trainning datasets. We randomly select 50 samples of our training datasets SHTE and the J.S. Bach's(JSB), respectively. Similarly, we randomly select 50 samples of our generated 8-bars long music pieces as the test dataset, covering different keys for testing. We compute the features mentioned above for these three datasets. Both absolute and relative measurements are discussed for these features mentioned above. The absolute measurements specially refer to the mean and the standard deviation, focusing on the characteristics within each dataset. On the other hand, the relative measurements emphasize the features across datasets, with the Kullback–Leibler divergence(KLD) and overlapped area(OA) being employed to evaluate the similarities between the generated music pieces and the two datasets. 
\begin{table*}[!h]
    \centering
    \caption{Comparison results for pitch feature of the training and generation data}
    \begin{tabular}{ccccccccccc}
        \toprule[1.5pt]
        \multirow{3}{*}{}&\multicolumn{2}{l}{\textbf{SHTE}} & \multicolumn{2}{l}{\textbf{Bach}} & \multicolumn{6}{l}{\textbf{Generated Data}}\\
        \cmidrule(r){2-3} \cmidrule(r){4-5} \cmidrule(r){6-11}
        \multirow{3}{*}{} &\multicolumn{2}{c}{Absolute measure} &\multicolumn{2}{c}{Absolute measure} & \multicolumn{2}{c}{Absolute measure} & \multicolumn{2}{c}{Inter-set(SHTE)} & \multicolumn{2}{c}{Inter-set(Bach)}\\
        \cmidrule(r){2-3} \cmidrule(r){4-5} \cmidrule(r){6-7} \cmidrule(r){8-9} \cmidrule(r){10-11}
        \multirow{3}{*}{} & Mean & STD & Mean & STD &Mean & STD & KLD & OA & KLD & OA\\
        \midrule[1.2pt]
        PC & 8.00 &1.79 &8.40 &2.06 &7.70 &1.27 &0.13 &0.67 &0.11 &0.71\\[2.5pt]
        PCH &- &- &- &- &- &- &0.16 &0.59 &0.21 &0.45 \\[2.5pt]
        DPR &0.96 &0.04 &0.93 &0.04 &0.86 &0.04 &0.02 &0.78 &0.03 &0.83\\[2.5pt]
        PR &11.9 &3.51 &11.7 &2.23 &9.4 &2.2 &0.05 & 0.74 &0.01 &0.82\\[2.5pt]
        PI &2.51 & 1.01& 1.92 &0.41 &2.33 &0.32 &0.11 &0.55 &0.04 &0.71\\[2.5pt]
        PCTM &- & -&- &- &- &- &0.09 &0.45 &0.19 & 0.28\\[2.5pt]
        \bottomrule[1.5pt]
    \end{tabular}
    \label{tab:absmetrics}
\end{table*}

Referring to the evaluation results, the analysis are as follows:
\begin{itemize}
    \item For the \textbf{PC} feature, SHTE and Bach have similar mean values, with SHTE at $8.00$ and Bach at $8.40$, though Bach’s standard deviation is slightly higher, suggesting a bit more variation in pitch class usage. The generated data has a lower mean ($7.70$) and a tighter spread ($STD=1.27$), which implies the generated sequences are centered around a set of pitch classes. About inter-set evaluation, the generated data has low KLD values for both SHTE ($0.13$) and Bach ($0.11$), indicating that its pitch class distribution aligns moderately well with both, but slightly more closely with Bach. The OA is $0.67$ with SHTE and $0.71$ with Bach, again suggesting the generated data’s pitch class distribution is slightly closer to Bach. 
    \item For \textbf{DPR}, SHTE and Bach have high diatonic pitch rates ($0.96$ and $0.93$, respectively), indicating a strong diatonic focus in both training sets. The generated data also has a high value but slightly lower mean DPR ($0.86$), implying that it includes more chromatic tones, which could add complexity and expressive range. Besides, the generated data has very low KLD values for DPR with both SHTE ($0.02$) and Bach ($0.03$), showing a strong resemblance to both, though marginally closer to SHTE. The overlap areas ($0.78$ with SHTE and $0.83$ with Bach) indicate that the generated data’s diatonic structure aligns well with both datasets. 
    \item About the \textbf{PR} feature, SHTE has a higher pitch range ($11.9$) compared to Bach ($11.7$), but both training datasets show substantial range variability. The generated data, with a mean of $9.4$ and lower standard deviation ($2.2$), suggests a more limited pitch range, potentially contributing to a more contained melodic structure. Meanwhile, the generated data has a lower KLD value with Bach ($0.01$) than with SHTE ($0.05$), indicating a pitch range distribution closer to Bach. Similarly, the OA is higher with Bach ($0.82$) than with SHTE ($0.74$), suggesting better alignment with Bach’s pitch range characteristics.
    \item Regarding \textbf{PI} feature, SHTE has a higher mean pitch interval ($2.51$) with more variability (STD $1.01$), indicating a wider range of interval sizes. Bach has a lower mean interval ($1.92$) and less variation, suggesting smoother transitions between notes. The generated data has a mean interval of $2.33$ with a smaller standard deviation ($0.32$), indicating it varies less than SHTE but more than Bach. The KLD values for PI are $0.11$ (SHTE) and $0.04$ (Bach), showing that the generated data’s pitch interval distribution aligns better with Bach. The overlap areas are $0.55$ with SHTE and $0.71$ with Bach, again indicating that the generated data’s intervals are closer to those in Bach.
    \item Besides, for the \textbf{PCH} and \textbf{PCTM} features,The KLD of PCH is lower for SHTE ($0.16$) than for Bach ($0.21$), meaning the pitch class histogram of the generated data is more aligned with SHTE. The OA values for PCH are $0.59$ (SHTE) and $0.45$ (Bach), also suggest that the generated data resembles SHTE more in pitch class distribution patterns. Similarly, KLD values for PCTM are higher with Bach ($0.19$) than with SHTE ($0.09$), and OA values are $0.45$ (SHTE) and $0.28$ (Bach), suggesting that the  pitch transitions of generated data more closely resemble SHTE. 
\end{itemize}
In summary, the generated data aligns more closely with Bach’s melodic characteristics while adopting some unique features from SHTE, such as a tighter pitch class spread and more distinct transition patterns, contributing to a balance of traditional structure and innovative variation. 

\textbf{B) Comparison with Baseline}

\begin{figure}[!h]
  \centering
  \includegraphics[width=1.0\columnwidth]{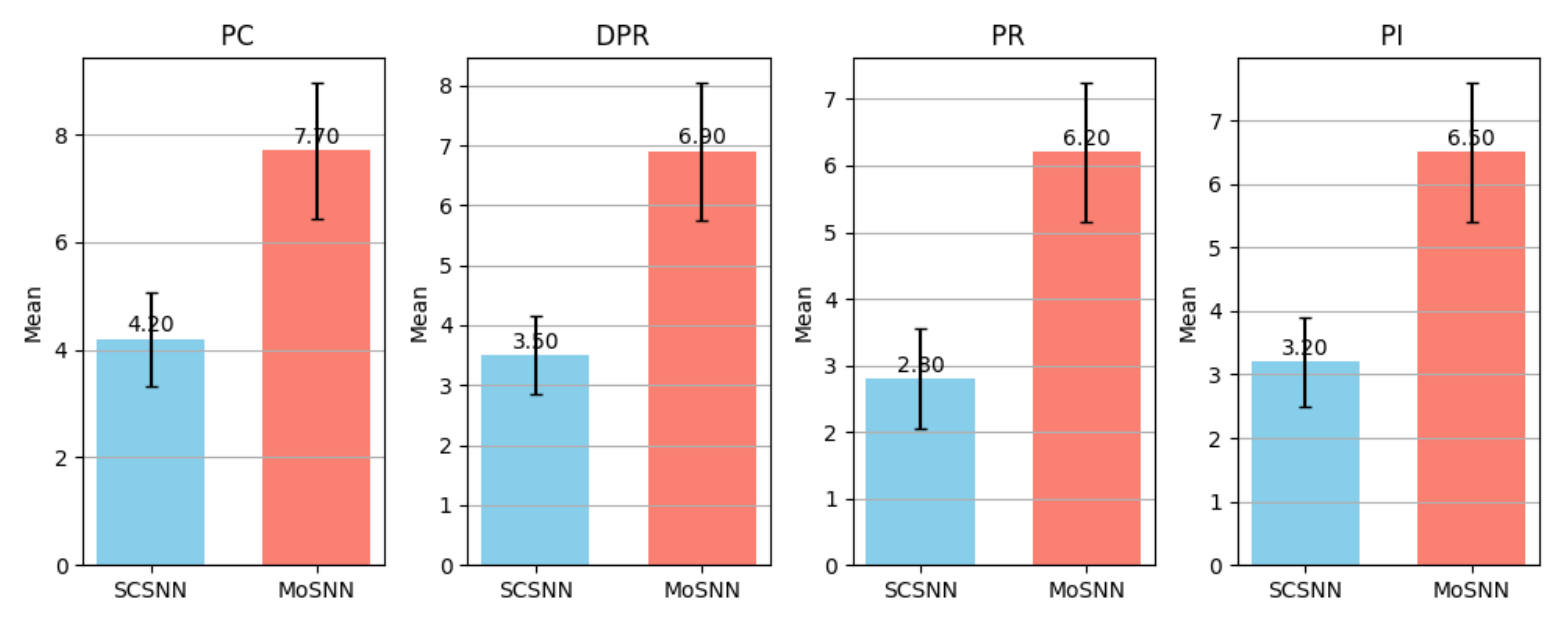}
  \caption{The comparison of the mean values between our model and the SCSNN baseline for the PC, DPR, PR, and PI metrics.}
  \label{fig:baseline-mean}
\end{figure}

Given the limited research on music generation using spiking neural networks, we focus on a single existing work based on brain-inspired SNN, referred to as \textbf{SCSNN} \cite{LQ2021}. Figure \ref{fig:baseline-mean} provides visual comparison of the mean values of four feature metrics (PC, DPR, PR, and PI) for the SCSNN baseline and our model. The error bars represent the standard deviation for each metric, indicating the variability within the results.  

The figure shows that, our model demonstrates a significantly higher mean \textbf{PC}, indicating that it generates music with a broader variety of pitches compared to SCSNN. The slightly higher standard deviation suggests that, while our model produces more varied music, it also exhibits greater variability in pitch diversity. For the \textbf{DPR} feature, our model achieves a higher rate than SCSNN(mean = 0.86 vs. 0.79) and reduced variability (std = 0.04 vs. 0.06), indicating that it generates music with pitches more closely aligned to the key's diatonic scale, thereby better preserving tonal consistency and producing harmonically coherent samples. Regarding \textbf{PR}, our model significantly outperforms SCSNN in range (mean = 9.4 vs. 5.9) but shows greater variability (std = 2.2 vs. 0.94), suggesting that it generates music with a broader tonal span. This increase in pitch range reflects greater melodic expressiveness and dynamic variety in the generated samples. The higher std could introduce unpredictability, it also highlights the model’s ability to explore a wider range of musical expressions, from simpler, narrower melodies to more expansive, complex ones. Finally, for \textbf{PI}, our model yields a larger average interval (mean = 2.33 vs. 1.68) with slightly lower variability (std = 0.32 vs. 0.38). 

\begin{figure}[!h]
  \centering
  \includegraphics[width=1.0\columnwidth]{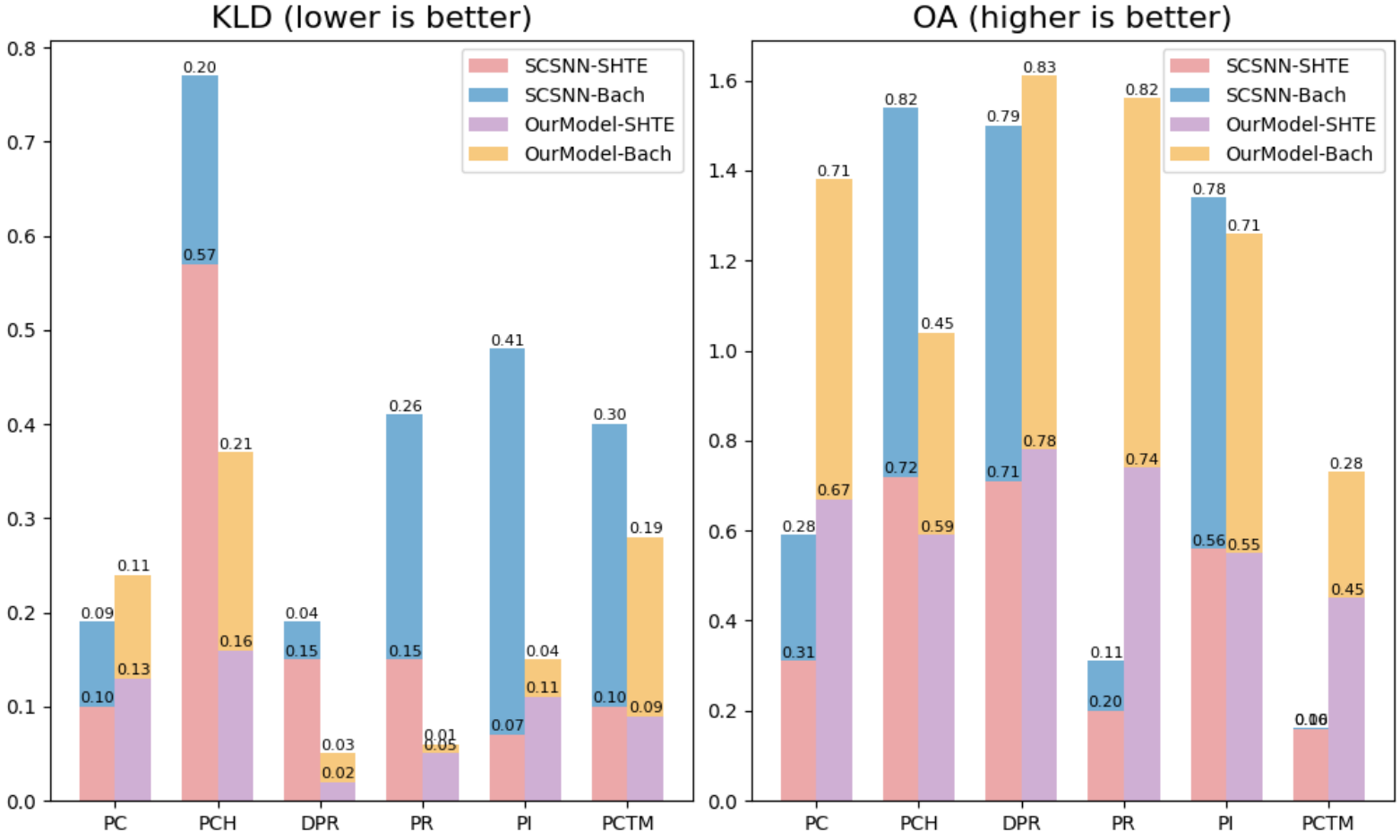}
  \caption{The performance comparison of KLDs and OAs for the features between our model and the baseline.}
  \label{fig:baseline}
\end{figure}

Besides, the comparison of two important metrics, KLD and OA, is presented in Fig.~\ref{fig:baseline}. Our model achieves lower KLD for PCH, PR, PI, and PCTM, with a particularly notable improvement in DPR, demonstrating better ability of the diatonic pitch structure. Specifically, for DPR, our model outperforms SCSNN with KLD values of 0.02 (SHTE) and 0.03 (Bach) compared to SCSNN’s 0.15 (SHTE) and 0.04 (Bach), indicating better adherence to tonal scales. The higher OA values for DPR further confirm that our model generates music that closely overlaps with the real data distributions, enhancing tonal coherence. For PR, our model shows significantly lower KLD and higher OA values on both datasets compared to SCSNN, indicating that our model’s pitch range better matches the actual data, leading to more realistic melodies. In PCH, our model performs better for SHTE with a KLD of 0.16 compared to SCSNN’s 0.57, while for Bach, both models show similar KLD values (0.21 for our model and 0.20 for SCSNN). However, SCSNN achieves higher OA for Bach, suggesting slightly better alignment for that dataset. Regarding PC, although both models have similar KLD values, our model achieves higher OA (0.67 for SHTE and 0.71 for Bach), indicating more consistent pitch generation. For PI, while SCSNN performs better for SHTE with a lower KLD (0.07 vs. 0.11), our model excels for Bach with a significantly lower KLD of 0.04 versus SCSNN’s 0.41. Combined with higher OA values, this result highlights our model’s improved accuracy in capturing pitch transitions. Lastly, in PCTM, our model outperforms SCSNN across both datasets, with lower KLD values (0.09 for SHTE and 0.19 for Bach) and higher OA, indicating better modeling of pitch class transitions.

 Overall, the results further indicate that the proposed model effectively captures both the tonality characteristics and the melody adaptability needed for generating diverse and musical content. The model’s ability to blend well-established tonal principles with novel musical expressions underscores its potential for generating musically meaningful compositions across various styles and datasets. This capability is essential for our future work on harmonic learning, emotion perception, and stylistic composition.

\section{Conclusion}
In this paper, we present a brain-inspired spiking neural network designed to learn the modes and keys of four-part music. Specifically, based on evolutionary neural circuits and collaborative areas, the model can learn the music theory of Western modes and keys, as well as the sequential relationships between notes. The trained model framework demonstrates the phenomena that closely resemble those of the well-known psychological model, the Krumhansl-Schmuckler model (KS model). Ultimately, the model is capable of generating the four-part music conditioned on different keys and modes. The evaluation results shows that the generated music pieces are musically coherent and reflect the characteristics of the specified modes and keys. This paper highlights the potential of brain-inspired spiking neural networks in advancing music generation technologies and provide a foundation for future research in this area.

Learning modes and keys is fundamental to advancing harmonic understanding and composition. In this research, we aim to explore brain-inspired models that are grounded in neuroscientific mechanisms and psychological theories to enhance our comprehension of music. By delving deeper into the cognitive processes that underpin musical perception and creativity, we hope to develop innovative approaches to music learning and generation. This interdisciplinary focus will not only inform the design of more effective neural network architectures but also contribute to a richer understanding of how humans interact with and create music.

As we move forward, our key focus will be on the relationship between harmonic learning and emotional generation. We believe that exploring how different harmonic structures can evoke various emotional responses will significantly enhance the capabilities of our models, allowing for the creation of music that resonates deeply with listeners. Through this work, we aim to bridge the gap between artificial intelligence, neuroscience, and music, providing new insights into this interdisciplinary domain. 

\section{Acknowledgments}
This work is supported by the Strategic Priority Research Program of the Chinese Academy of Sciences (Grant No.XDB1010302).

\subsection{Author contributions}
Q.L designed the study, implemented the code, conducted the experiments, and drafted the manuscript. Y.Z. supervised the project and contributed to the manuscript writing. MHR.T. assisted with code maintenance.

\subsection{Declaration of interests}
The authors declare that they have no competing interests.

\bibliographystyle{cas-model2-names}
\bibliography{mybibfile}

\end{document}